\documentclass[preprint,showpacs,preprintnumbers,amsmath,amssymb,natbib]{revtex4}

\usepackage{graphicx}
\usepackage{epsfig}
\usepackage{dcolumn}
\usepackage{bm}

\def\be{\begin{equation}}
\def\ee{\end{equation}}
\def\ba{\begin{eqnarray}}
\def\ea{\end{eqnarray}}

\begin{document}

\title{Entangled States and the Gravitational Quantum Well}

\author{Rui Alves\footnote{E-mail: up201106579@fc.up.pt}}
\affiliation{Departamento de F\'isica e Astronomia, Faculdade de Ci\^encias da Universidade do Porto, Rua do Campo Alegre, 687, 4169-007 Porto, Portugal}

\author{Catarina Bastos\footnote{E-mail: catarina.bastos@tecnico.ulisboa.pt}}
\affiliation{GoLP/Instituto de Plasmas e Fus\~ao Nuclear, Instituto Superior T\'ecnico, Universidade de Lisboa, Avenida Rovisco Pais 1, 1049-001 Lisboa, Portugal}

\author{Orfeu Bertolami\footnote{Also at Centro de F\'isica do Porto, Rua do Campo Alegre, 687, 4169-007 Porto, Portugal. \linebreak E-mail: orfeu.bertolami@fc.up.pt}}
\affiliation{Departamento de F\'isica e Astronomia, Faculdade de Ci\^encias da Universidade do Porto, Rua do Campo Alegre, 687, 4169-007 Porto, Portugal}

\date{\today}

\begin{abstract}
We study the continuous variable entanglement of a system of two particles under the influence of Earth's gravitational field. We determine a phase-space description of this bipartite system by calculating its Wigner function and verify its entanglement by applying a generalization of the PPT criterion for non-Gaussian states. We also examine the influence of gravity on an idealized entanglement protocol to be shared between stations at different potentials based on the correlation of states of the gravitational quantum well.
\end{abstract}

\maketitle

\section{Introduction}

The rapid development of quantum technologies, such as quantum cryptography \cite{Ekert, Gisin} and quantum teleportation \cite{Bennett}, has been one of the main propellers for the study of quantum entanglement in recent years \cite{Horodecki, Adesso}. In fact, it is quite relevant to study the behaviour of entangled particles separated by large distances, as this will be crucial for the creation of global quantum communication systems. It seems consensual that the next step in this direction is to develop new space-based experiments. For instance, a satellite-based distribution of entangled keys would allow not only for the development of a worldwide network of quantum communication, but also for the improvement of tests of quantum mechanics.

Indeed, recently, a team of researchers has shown that observers separated by 144 km can share a quantum cryptographic key \cite{Ursin} by exploiting the randomness and strong quantum correlations inherent to quantum entanglement. As a result of this, several proposals have been made to study the entanglement between a station on Earth's surface and a second station at orbit \cite{Rideout, Bruschi}.

Inspired by these developments, we study here the effect of Earth's gravitational field on entangled neutron states. Prior work on this topic can be found in Refs. \cite{Li:2012zt, Hasegawa}. We analyze the effect of having entangled particles at different values of the gravitational potential and discuss how this could affect an entanglement protocol between an observer in a station on the surface of the Earth and a second observer in a station in a Low-Earth Orbit (LEO).

Of course, gravity, being the feeblest of all known forces, is not expected to upset a given entanglement protocol. Nevertheless, the importance of understanding gravity in a quantum mechanical context impels us to pursue this line of theoretical inquiry and examine the effect of gravity on entangled quantum states.

The problem of a particle under the influence of a constant gravitational field is well known in quantum mechanics \cite{Vallee}. Consider a particle of mass M in a gravitational field ${\bf g}=-g{\bf e_x}$. When an horizontal mirror is placed at $x=0$, a gravitational quantum well (GQW) is established \cite{06A}, and the system is described by a potential well of the form $V(x) = Mgx$, for $x\geq0$. The solutions to the eigenvalue equation, $H\psi_n=E_n\psi_n$, can be expressed in terms of the Airy function of first type,
\begin{equation} \label{eq:airy-def}
\psi_n(x) = A_n\text{Ai}\left(\frac{x-x_n}{x_0}\right),
\end{equation}
while the energy eigenvalues are determined by the roots of the Airy function, $\alpha_n$, with $n=1,2,...,$
\begin{equation} \label{eq:airy-energy-def}
E_n=-\left(\frac{Mg^2\hbar^2}{2}\right)^{1/3}\alpha_n.
\end{equation}
Here, $x_0 \equiv \left( \hbar^2/2M^2g \right)^{1/3}$ is a scaling factor, $A_n$ is the normalization constant of the {\it n}-th level, and $x_n = E_n/Mg$ corresponds to the maximum height classically allowed for a particle with energy $E_n$.

The probability of finding the particle is non-vanishing for all values of $x>0$. However, it will have a maximum value at the classical turning point $x_n$. Above this height, the probability of finding the particle decays exponentially.

This quantum system was built experimentally by submitting a beam of ultra-cold neutrons to Earth's gravitational well that bounces on a horizontal mirror \cite{Nesvizhevsky}. Ultra-cold neutrons are fundamental for this experiment, as they are less likely to be affected by the electromagnetic interaction. In simple terms, the experiment runs as follows: a scatterer/absorber is placed above the horizontal mirror, forming a slit, and the neutron transmission through this slit is measured. If the height of the the scatterer/absorber is larger than the classical turning point for a given quantum state, the neutrons pass through the slit without loss. As the size of the slit decreases, the probability of neutron loss increases until the slit size reaches $x_n$ and the apparatus stops being transparent to neutrons in the {\it n}-th quantum state. This procedure allows also for a criterion for the transition from quantum to classical behaviour \cite{OB2}.

In this work, we study the influence of gravity on entanglement of states. We consider entangled GQW states that clearly depend on a continuous variable. Gaussian states are the most studied continuous variable states in quantum information theory \cite{Adesso}. They have been shown to be useful to construct entanglement protocols and are important testbeds for investigating quantum correlations. These states are usually described by their covariance matrices, which are built from the second statistical moments of the states and encode all the information about them.

The elements of a covariance matrix, $\bm{\sigma} = \left( \sigma_{ij} \right)$, can be determined by the relations
\begin{equation} \label{eq:covariance}
\sigma_{ij} = \langle \hat{R}_i\hat{R}_j + \hat{R}_j\hat{R}_i \rangle_\rho - 2 \langle \hat{R}_i \rangle_\rho \langle \hat{R}_j \rangle_\rho,
\end{equation}
where $\bm{\hat{R}} = (\hat{q}_1, \hat{p}_q, ..., \hat{q}_N, \hat{p}_N)^T$ is a vector of the quadrature operators, and $\langle \hat{O} \rangle_\rho \equiv \text{Tr}[ \rho \hat{O} ]$ denotes the mean of the operator $\hat{O}$ evaluated with a density matrix $\rho$.

Since quantum Gaussian states can be completely described by their second moments, these covariance matrices are the central elements of the separability or entanglement criteria. The positivity of the partial transpose (PPT) of the covariance matrix is shown to be a necessary and sufficient condition for the separability of a bipartite Gaussian state \cite{Simon}.

In what follows, we shall see that the states of the GQW system are not Gaussian and, thus, an extension of the criteria for the separability of states has to be considered.

This paper is organized as follows: In section II, we present the mathematical tools to study entanglement in a bipartite state of the GQW. In section III, we study the effects of considering two particles at different potentials. Finally, in section IV, we discuss our results and present our conclusions.

\section{Entanglement in the Gravitational Quantum Well}

We start by considering a bipartite state of the form
\begin{equation} \label{eq:state}
| \psi^{+} \rangle = \frac{1}{\sqrt{2}} \left( | nm \rangle + | mn \rangle \right),
\end{equation}
where $|n\rangle$ denotes the {\it n}-th level of the GQW, that is, the state of a neutron in a gravitational field $\bf{g}$ with energy $E_n$ and wavefunction $\langle x | n \rangle = \psi_n(x)$. When studying bipartite systems we use $|nm\rangle = |n\rangle_{A} \otimes |m\rangle_{B}$ to denote a two-particle system with a particle in the energy level {\it n} in subsystem A and a particle in the energy level {\it m} in subsystem B.

To study continuous variable entanglement in this bipartite system, we proceed by determining its Wigner function, which in case of Gaussian states would be a Gaussian function. First, we consider the density matrix for state of Eq. (\ref{eq:state}):
\begin{equation}
\hat{\rho} = |\psi^+\rangle\langle\psi^+| = \frac{1}{2}\left( | nm \rangle\langle nm| + | nm \rangle\langle mn| + | mn \rangle\langle nm| +| mn \rangle\langle mn| \right).
\end{equation}
We now proceed by applying this density matrix to the Wigner function
\begin{equation} \label{eq:wigner-def}
W(x_A, x_B;p_A, p_B) = \frac{1}{\pi^2}\iint \langle x_A + q_A, x_B + q_B | \hat{\rho} | x_A-q_A, x_B-q_B \rangle e^{2i(p_Aq_A+p_Bq_B)}dq_Adq_B.
\end{equation}

For the computation of the Wigner function, the following relationships for Airy functions are quite useful \cite{Vallee}:
\begin{align} \label{eq:wigner-integral1}
\int_{-\infty}^{+\infty} \text{Ai}\left( \frac{x-x_n+q}{x_0} \right)\text{Ai}\left( \frac{x-x_n-q}{x_0} \right)e^{2ipq}dq =  \nonumber \\ 
= \frac{x_0}{2^{1/3}}\text{Ai}\left( 2^{2/3}\left( \frac{x-x_n}{x_0} + x_0^2p^2 \right) \right),
\end{align}
\begin{align} \label{eq:wigner-integral2}
\int_{-\infty}^{+\infty} \text{Ai}\left( \frac{x-x_n+q}{x_0} \right)\text{Ai}\left( \frac{x-x_m-q}{x_0} \right)e^{2ipq}dq = \nonumber \\
 = \frac{x_0}{2^{1/3}} \text{Ai}\left(\frac{2x-x_n-x_m}{2^{1/3}x_0} + 2^{2/3}x_0^2p^2 \right)e^{i\left(x_n - x_m\right)p}.
\end{align}

Using these relationships, it is straightforward to show that the Wigner function for the bipartite state of Eq. (\ref{eq:state}) takes the form:
\begin{align} \label{eq:wigner}
& W(x_A, x_B;p_A, p_B) = \frac{A_n^2 A_m^2}{2 \pi^2}\frac{x_0^2}{2^{2/3}} \bigg\{\text{Ai}\left( 2^{2/3} \left( \frac{x_A-x_n}{x_0} + x_0^2p_A^2 \right) \right) \text{Ai}\left( 2^{2/3} \left( \frac{x_B-x_m}{x_0} + x_0^2p_B^2 \right) \right) \nonumber \\
 & \quad + \text{Ai}\left( \frac{2x_A-x_n-x_m}{2^{1/3}x_0} + 2^{2/3}x_0^2p_A^2 \right) \text{Ai}\left( \frac{2x_B-x_n-x_m}{2^{1/3}x_0} + 2^{2/3}x_0^2p_B^2 \right) e^{i\left( x_n-x_m \right)p_A}e^{i\left( x_m-x_n \right)p_B} \nonumber \\
 & \quad + \text{Ai}\left( \frac{2x_A-x_n-x_m}{2^{1/3}x_0} + 2^{2/3}x_0^2p_A^2 \right) \text{Ai}\left( \frac{2x_B-x_n-x_m}{2^{1/3}x_0} + 2^{2/3}x_0^2p_B^2 \right) e^{i\left( x_m-x_n \right)p_A}e^{i\left( x_n-x_m \right)p_B} \nonumber \\
 & \quad + \text{Ai}\left( 2^{2/3} \left( \frac{x_A-x_m}{x_0} + x_0^2p_A^2 \right) \right) \text{Ai}\left( 2^{2/3} \left( \frac{x_B-x_n}{x_0} + x_0^2p_B^2 \right) \right) \bigg\}.
\end{align}

It is clear that the system we are considering is not Gaussian; however, having a phase-space description of the system through the Wigner function, we are in condition to build the matrices of moments and to carry out continuous variable entanglement tests.

The {\it n}-th statistical moment of an operator $\hat{O}$ can be extracted from the Wigner function \cite{Deesuwan}:
\begin{equation}
\langle O^n \rangle_\rho = \int W(x_A, x_B;p_A, p_B) O^n(x_A, x_B;p_A, p_B) dx_Adx_Bdp_Adp_B,
\end{equation}
where the integrals are calculated over the allowed region of the phase space. For  the GQW, this corresponds to the range $x_i \in [0, +\infty[$ in the position variables and $p_i \in \mathbb{R}$ in the momentum variables ($i = \text{A, B}$).

We are interested in determining moments up to second order in position and momentum. To achieve this, we see from Eq. (\ref{eq:wigner}) that we have to calculate two types of integrals. We call integrals of type I those of the form
\begin{equation} \label{eq:type1}
\int_{0}^{+\infty}\int_{-\infty}^{+\infty} O(x, p) \text{Ai}\left( 2^{2/3} \left( \frac{x-x_n}{x_0} + x_0^2p^2 \right) \right) dp dx,
\end{equation}
and type II those of the form
\begin{equation} \label{eq:type2}
\int_{0}^{+\infty}\int_{-\infty}^{+\infty} O(x, p) \text{Ai}\left( \frac{2x-x_n-x_m}{2^{1/3}x_0} + 2^{2/3}x_0^2p^2 \right) e^{i\left( x_n-x_m \right)p}dp dx,
\end{equation}
where $O(x, p)$ represents the combination of variables corresponding to the expected statistical moment.

Most of these integrals are not standard, and thus require a considerable amount of manipulations to be computed. Those of the form of Eq. (\ref{eq:type2}) are particularly challenging to evaluate. However, Airy functions possess many algebraic and cyclic properties that can be exploited in order to calculate these integrals. Ref. \cite{Vallee} is an excellent resource for this task.

Performing each of the these calculations, we arrive at the results presented in Table \ref{table:integrals}.

\begin{table}[h]
\def\arraystretch{1.2}
\begin{tabular}{ c c c }
\hline
\hline
 $O(x,p)$ & Type I & Type II \\ 
 \hline
 $1$ &  $\frac{2^{1/3}\pi}{x_0A_n^2}$ & $0$ \\ 
 $x$ & $-\frac{2^{4/3}\pi}{3A_n^2}\alpha_n$ & $-\frac{2^{4/3}\pi}{A_nA_m}\left(\frac{1}{\alpha_m - \alpha_n} \right)^2$ \\
 $x^2$ & $\frac{2^{1/3}8\pi x_0}{15A_n^2}\alpha_n^2$ & $-\frac{2^{1/3}24\pi x_0}{A_nA_m}\left( \frac{1}{\alpha_m - \alpha_n} \right)^4$ \\
 $p$ & $0$ & $-\frac{2\pi}{x_0^2 A_n A_m}\left( \frac{1}{\alpha_m-\alpha_n} \right)$ \\
 $p^2$ & $-\frac{2^{4/3}\pi}{3x_0^3A_n^2}\alpha_n^2$ & $-\frac{2^{2/3}4\pi}{x_0^3A_nA_m} \left( \frac{1}{\alpha_m-\alpha_n}\right)^2$ \\
 $xp$ & $0$ & $\frac{12\pi}{x_0A_nA_m} \left( \frac{1}{\alpha_m-\alpha_n}\right)^3$ \\
\hline
\hline
\end{tabular}
\caption{Results for the integrals of Eqs. (\ref{eq:type1}) and (\ref{eq:type2}) with various combinations $O(x,p)$ of position and momentum variables.}
\label{table:integrals}
\end{table}

From here, we build the statistical moments by matching the results of Table \ref{table:integrals} in the way required by the Wigner function. From Eq. (\ref{eq:wigner}), we see that the function is composed of four terms, each of which requires the calculation of two integrals: one for subsystem A and another for subsystem B. Two of these terms involve only integrals of type I, while the other two involve solely integrals of type II.

As an example of calculation, we use the results for $O(x, p)=1$ to verify that
\begin{eqnarray}
\int W(x_A, x_B;p_A, p_B) dx_Adx_Bdp_Adp_B &=& \frac{A_n^2 A_m^2}{2 \pi^2}\frac{x_0^2}{2^{2/3}} \bigg\{ \frac{2^{1/3}\pi}{x_0A_n^2}\frac{2^{1/3}\pi}{x_0A_m^2} + \frac{2^{1/3}\pi}{x_0A_m^2}\frac{2^{1/3}\pi}{x_0A_n^2} \bigg\} \nonumber \\
&=& 1,
\end{eqnarray} 
which is one of the main properties of any Wigner function. The only terms contributing to this calculation are those of integrals of type I, since those of type II vanish.

Now we use Eq. (\ref{eq:covariance}) to build the covariance matrix of the bipartite state, which can be shown to take the form:
\begin{equation} \label{eq:covariance-state}
\bm{\sigma} = 
\begin{pmatrix}
x_0^2\alpha & 0 & x_0^2 \beta & 0 \\
0 & \gamma/x_0^2 & 0 & \delta/x_0^2 \\
x_0^2\beta & 0 & x_0^2 \alpha & 0 \\
0 & \delta/x_0^2 & 0 & \gamma/x_0^2 \\
 \end{pmatrix}, 
 \end{equation}
where
\begin{eqnarray}
\alpha &=& \frac{14}{45}\left( \alpha_n^2 + \alpha_m^2 \right) - \frac{4}{9}\alpha_n\alpha_m \\
\beta &=& \frac{8}{\left(\alpha_m - \alpha_n\right)^2} - \frac{2}{9}\left( \alpha_m-\alpha_n \right)^2 \\
\gamma &=& -4\left( \alpha_m + \alpha_n \right) \\
\delta &=& -2 \left( \frac{2^{2/3}}{\alpha_m-\alpha_n} \right)^2
\end{eqnarray}

As it has already been referred, to study Gaussian entanglement we would simply have to apply a Gaussian test of entanglement to this matrix. A necessary and sufficient condition for separability of Gaussian states is given by the Peres-Horodecki criterion \cite{Simon}. When applied to matrix of Eq. (\ref{eq:covariance-state}), the criterion takes the form $\Delta \geq 0$, where
\begin{equation} \label{eq:peres-horodecki}
\Delta = (\alpha\gamma)^2 - (\alpha\delta)^2 - (\beta\gamma)^2 + \frac{1}{16} - \frac{|\beta\delta|}{2} + (\beta\delta)^2 - \frac{\alpha\gamma}{2}.
\end{equation}

\begin{figure}[h]
	\centering
	\includegraphics[width=0.7\textwidth]{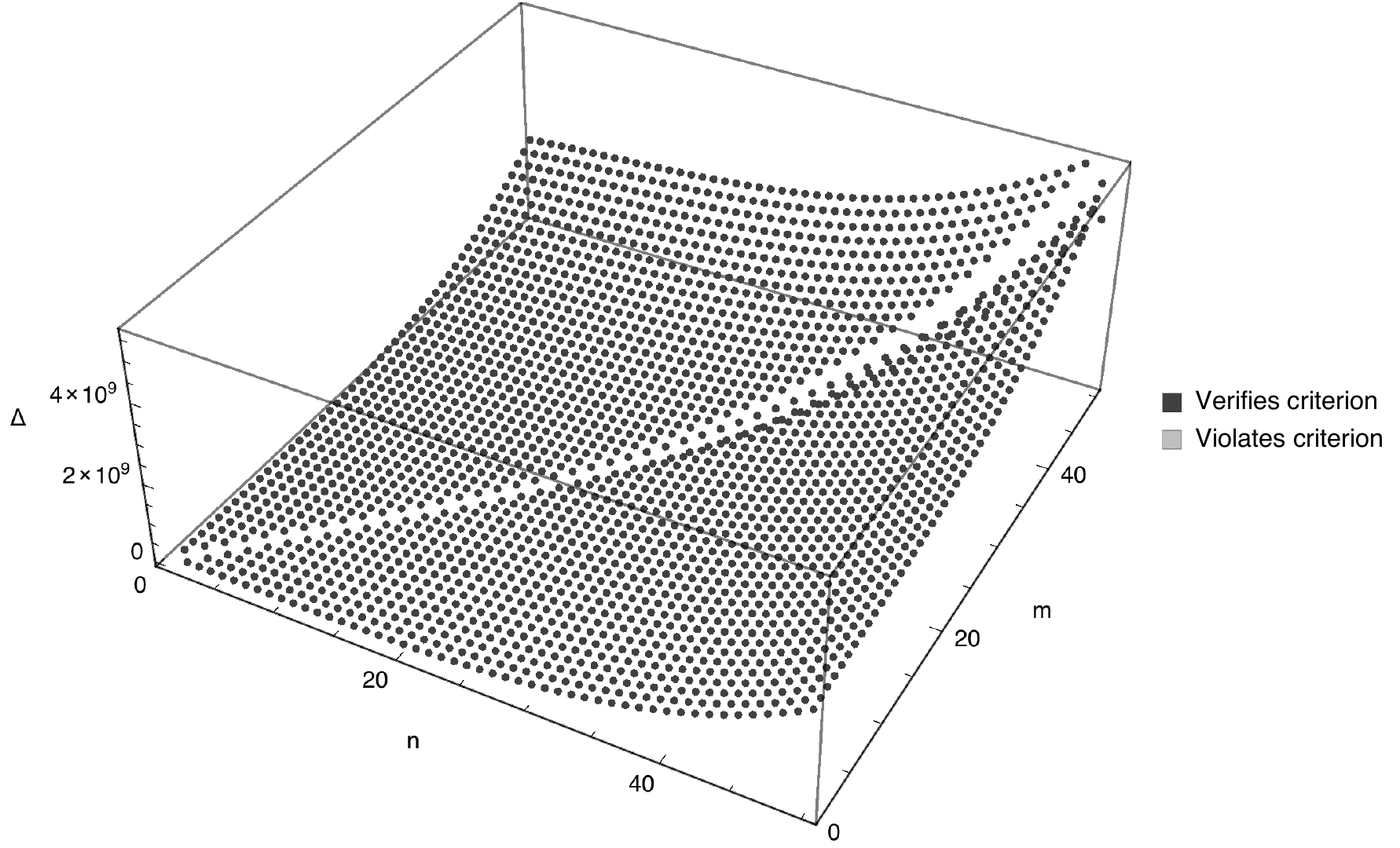}
	\caption{Numerical results for the application of the Peres-Horodecki criterion for various combinations of {\it n} and {\it m}, up to the energy level $50$.}
	\label{fig:gaussian-criterion}
\end{figure}

It is easy to show, using the numerical values of the zeros $\alpha_n$, that all GQW states satisfy this condition and, thus, there is no evidence of entanglement at this level. In fact, considering the two lowest energy levels, $n=1$ and $m=2$, the criterion yields $\Delta~=~4573.31~>~0$, which implies the states are separable. Considering higher energy levels, the value in the left-hand side becomes larger and the separability criterion is never broken. Fig. \ref{fig:gaussian-criterion} shows the numerical results for the application of the criterion for different energy levels.

However, the Peres-Horodecki criterion is only a necessary and sufficient condition of separability when applied to Gaussian states. For non-Gaussian states, second order criteria may fail to reveal entanglement \cite{Gomes}. Thus, genuine entanglement of non-Gaussian states is only revealed through application of criteria involving higher-order moments.

To achieve this, we follow the construction presented in Ref. \cite{Miranowicz}, where it is developed a generalization of the Positive Partial Transposition (PPT) criterion for continuous variable systems based on the matrices of moments. The PPT criterion states that a separable state remains positive under partial transposition, and, therefore, a Non-positive-Partial-Transposition (NPT) state must be entangled.

We start by defining adimensional operators $a$ and $a^\dagger$ for subsystem A, and $b$ and $b^\dagger$ for subsystem B, such that

\begin{eqnarray}
a &=& \frac{1}{\sqrt{2}} \left( \frac{x_A}{x_0} + ix_0p_A \right), \quad a^\dagger = \frac{1}{\sqrt{2}} \left( \frac{x_A}{x_0} - ix_0p_A \right), \label{eq:oper-a}\\
b &=& \frac{1}{\sqrt{2}} \left( \frac{x_B}{x_0} + ix_0p_B \right), \quad b^\dagger = \frac{1}{\sqrt{2}} \left( \frac{x_B}{x_0} - ix_0p_B \right). \label{eq:oper-b}
\end{eqnarray}

We build now a suitable matrix of moments $M_{f}(\rho) = [M_{ij}] = [\langle f_i^\dagger f_j \rangle]$ that forms the basis for the criterion. Let $\rho^\Gamma$ denote partial transposition of the state $\rho$ with respect to subsystem B. Then, the criterion reads as follows: {\it a bipartite state $\rho$ is NPT if and only if there exists $f$ such that} $\text{det\,} M_f(\rho^\Gamma)$ {\it is negative} \cite{Miranowicz}. Moreover, it can be shown that if the class $f$ of operators has a tensor product structure, $\tilde{f}=f^A \otimes f^B$, then $M_{\tilde{f}}(\rho^\Gamma) = (M_{\tilde{f}}(\rho))^\Gamma$.

We choose $\tilde{f}=(1, a) \otimes (1, b) = (1, a, b, ab)$ and the corresponding matrix of moments becomes

\begin{equation} \label{eq:ppt-matrix}
M_{\tilde{f}}(\rho) = 
\begin{pmatrix}
1 & \langle a \rangle & \langle b \rangle & \langle ab \rangle \\
\langle a^\dagger \rangle & \langle a^\dagger a \rangle & \langle a^\dagger b \rangle & \langle a^\dagger ab \rangle \\
\langle b^\dagger \rangle & \langle ab^\dagger \rangle & \langle b^\dagger b \rangle & \langle a b^\dagger b \rangle \\
\langle a^\dagger b^\dagger \rangle & \langle a^\dagger a b^\dagger \rangle & \langle a^\dagger b^\dagger b \rangle & \langle a^\dagger a b^\dagger b \rangle \\
 \end{pmatrix}.
 \end{equation}
 
The next step consists in rewriting the statistical moments of matrix Eq. (\ref{eq:ppt-matrix}) in terms of the statistical moments of the momentum and position variables. Performing this allows us to determine the entries of $(M_{\tilde{f}}(\rho))^\Gamma$ by applying the results from Table \ref{table:integrals}.

\begin{figure}[h]
	\centering
	\includegraphics[width=0.7\textwidth]{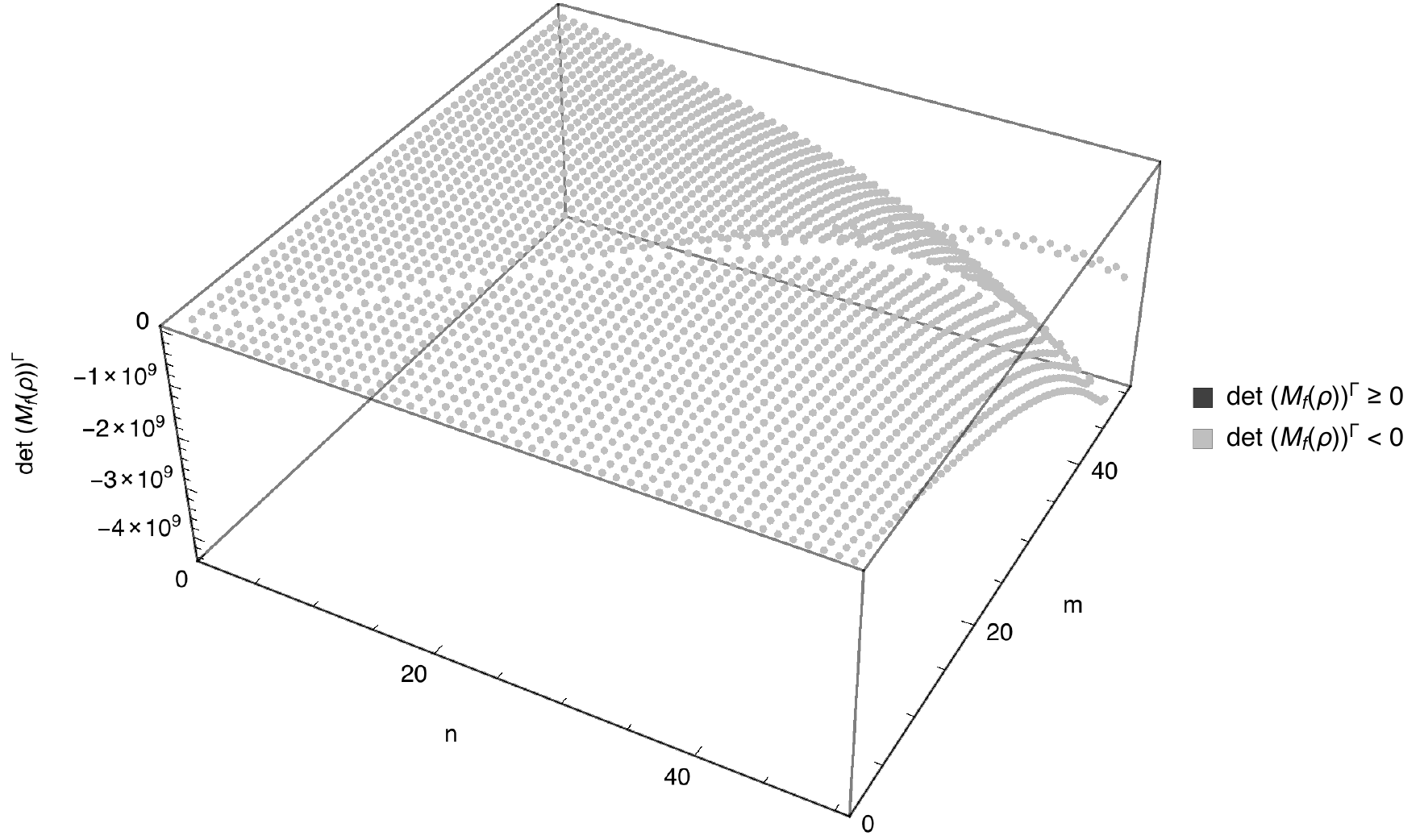}
	\caption{Numerical results for the application of the non-Gaussian generalization of the PPT criterion for various combinations of {\it n} and {\it m}, up to the energy level $50$.}
	\label{fig:non-gaussian-criterion}
\end{figure}

Substituting the numerical values for the zeros of Airy functions $\alpha_n$, it can be shown that any combination of GQW states $n$ and $m$ satisfies the condition $\text{det\,} (M_{\tilde{f}}(\rho))^\Gamma < 0$. These results are plotted in Fig. \ref{fig:non-gaussian-criterion}. As an example, if we consider $n=1$ and $m=2$, $\text{det\,} (M_{\tilde{f}}(\rho))^\Gamma = -0.588169$. For higher energy levels, this value remains negative. Hence, we conclude that state described by Eq. (\ref{eq:state}) is NPT and the system is entangled.

\section{Moving in the Gravitational Field}

We aim now to study the behavior of this entangled system when one of its parts moves vertically in the gravitational field. To achieve this, we consider that the particle in subsystem B is displaced to a height $H$ relative to that of subsystem A and, at this new position, the particle feels a gravitational field $g'<g$.

Looking back at the definitions of the GQW in Eqs. (\ref{eq:airy-def}) and (\ref{eq:airy-energy-def}), we see that the new subsystem B is dependent on the constants $x_0' = \left( \hbar^2/2M^2g' \right)^{1/3}$ and $x_n' = -x_0' \alpha_n$, and the new energy levels are given by

\begin{equation}
	E_n'=-\left(\frac{Mg'^2\hbar^2}{2}\right)^{1/3}\alpha_n = -Mg'x_0'\alpha_n.
\end{equation}

We consider again a state of the form
\begin{equation}
	| \psi^{+} \rangle = \frac{1}{\sqrt{2}} \left( | nm' \rangle + | mn' \rangle \right),
\end{equation}
where $|n'\rangle$ is the {\it n}-th energy level of the particle in potential $g'$. 

The new Wigner functions is, thus, dependent on both $x_0$ and $x_0'$, respectively due to the positions of particle A and particle B. It is also important to note that new normalization factors $A_n'$ must be included in the Wigner description. However, it is straightforward to see from the results of Table \ref{table:integrals} that these normalization factors are easily factored out when we compute any type of statistical moment.

The statistical moments are obtained in the same fashion as those of the previous section: we can, again, distinguish between the two types of integrals and we build the moments by pairing the results of Table \ref{table:integrals} in the way required by the Wigner function. The only difference is that now all the results involving subsystem B are affected by $x_0'$ instead of $x_0$.

We want to study how the criterion presented in the previous section is affected by this change in the gravitational field. For this purpose we build the adimensional operators of Eqs. (\ref{eq:oper-a}) and (\ref{eq:oper-b}) with the new statistical moments in B. Since these moments on the position and momentum variables now depend on $x_0'$, all moments in $b$ and $b^\dagger$ become affected by the ratio
\begin{equation}
	\frac{x_0'}{x_0} = \frac{\left( \hbar^2/2M^2g' \right)^{1/3}}{\left( \hbar^2/2M^2g \right)^{1/3}} = \left( \frac{g}{g'} \right)^{1/3}.
\end{equation}

\begin{figure}[h]
	\centering
	\includegraphics[width=0.7\textwidth]{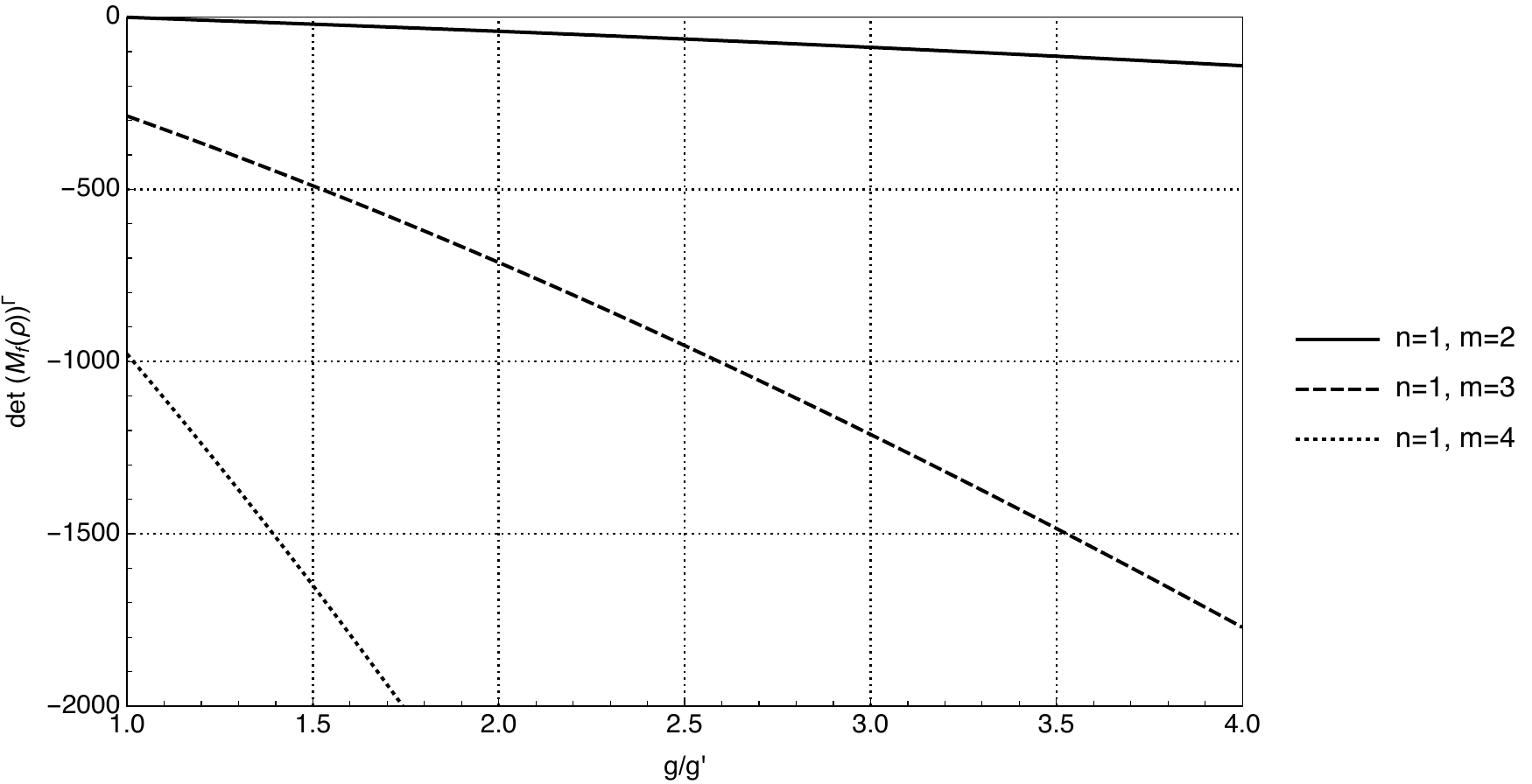}
	\caption{Numerical results for the application of the non-Gaussian generalization of the PPT criterion for different values of $g/g'$ on three combinations of $n$ and $m$}
	\label{fig:variable-g}
\end{figure}

\begin{figure}[h]
	\centering
	\includegraphics[width=1\textwidth]{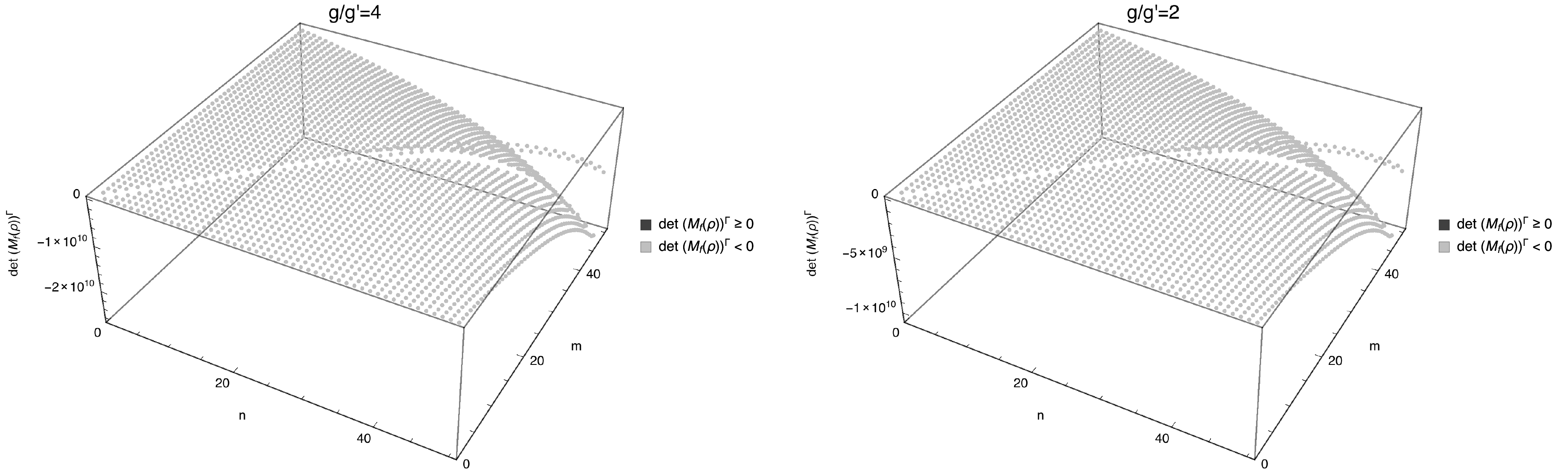}
	\caption{Numerical results for the application of the non-Gaussian generalization of the PPT criterion for two different values of the ratio $g/g'$ and multiple combinations of $n$ and $m$, up to the energy level $50$.}
	\label{fig:variable-g-nm}
\end{figure}

\begin{figure}[h]
	\centering
	\includegraphics[width=0.70\textwidth]{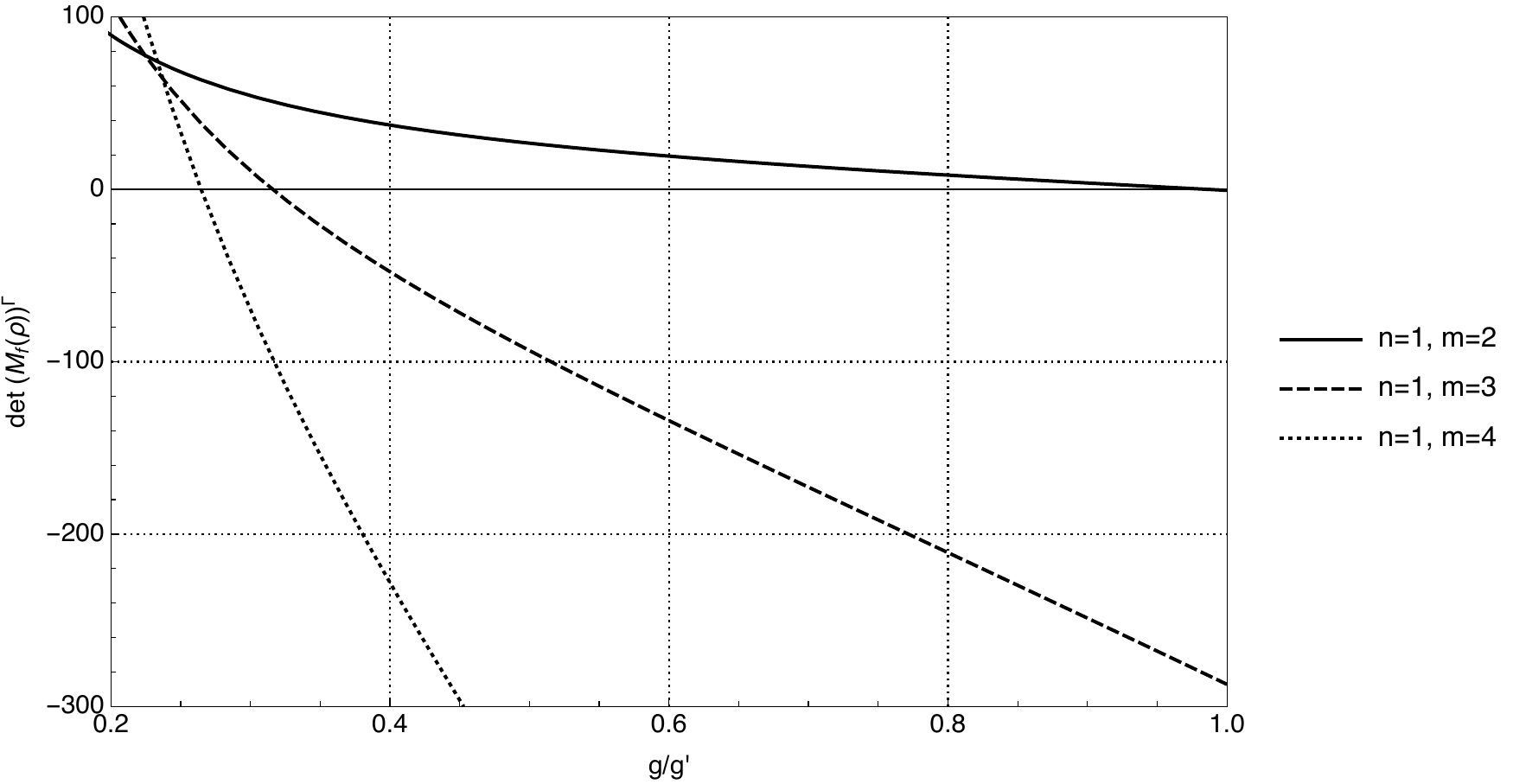}
	\caption{Numerical results for the application of the non-Gaussian generalization of the PPT criterion for different values of $g/g'<1$ on three combinations of $n$ and $m$}
	\label{fig:variable-g-pos}
\end{figure}

\begin{figure}[h]
	\centering
	\includegraphics[width=1\textwidth]{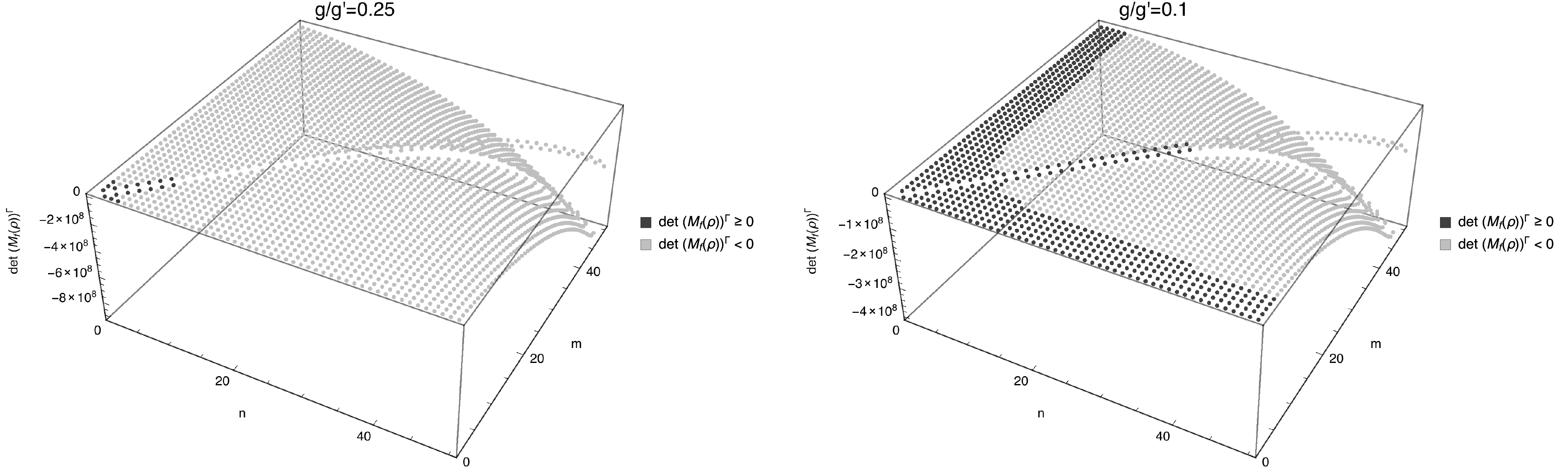}
	\caption{Numerical results for the application of the non-Gaussian generalization of the PPT criterion for two different values of the ratio $g/g'<1$ and multiple combinations of $n$ and $m$, up to the energy level $50$.}
	\label{fig:variable-g-nm-pos}
\end{figure}

We now build the matrix of moments of Eq. (\ref{eq:ppt-matrix}) and apply the PPT criterion for different values of $g/g'$. The results are shown in Figures \ref{fig:variable-g}-\ref{fig:variable-g-nm-pos}.

It is clear from the results of Figures \ref{fig:variable-g} and \ref{fig:variable-g-nm} that if the particle B feels a gravitational field $g'$ weaker than $g$, then the determinant of the matrix of moments remains negative and, thus, the system remains NPT. We expect, therefore, that moving the particle upwards in the gravitational field does not break the entanglement that we have examined in the previous section for the case where the two particles are at the same height.

Notice that if we could achieve a scenario where $g'>g$, the results would not be as simple. As we can see from Figures \ref{fig:variable-g-pos} and \ref{fig:variable-g-nm-pos}, for extreme cases of $g/g'<1$ the determinant becomes positive and the generalization of PPT criterion fails to reveal entanglement. However, due to the nature of the criterion, it is unclear if the entanglement is indeed broken by the stronger gravitational field or if it is simply a limitation imposed by our choice for the class of operators $\tilde{f}$.

Following this, it is clear that if we could devise a system similar to that of Ref. \cite{Ursin}, which was used to verify the presence of entanglement between two observers separated by 144 km, the entanglement of the states would not be disrupted by gravity.

We can think of a system composed of two parts: station A, located at the surface of the Earth; and station B, a platform at height $H$. Station A would be responsible for producing the entangled bipartite state and sending one of the particles to the other station. Ideally, station B would be an orbiting platform placed in Low-Earth Orbit (LEO), allowing for a separation of hundreds of kilometers between the two stations. As we have seen, moving one of the particles to station B, which is in a weaker region of the gravitational field, should not break the entanglement of the bipartite system. More details on the proposals for satellite-based tests of quantum entanglement and, in particular, for platforms at LEO can be found in Refs. \cite{Rideout, Bruschi}.

\section{Conclusions}

In this work we have studied the effect of gravity on the entanglement of states. We have built a phase-space description of a bipartite system of two particles in the GQW by calculating its corresponding Wigner function. We have shown that this Wigner description reveals a non-Gaussian state.

We have also shown that Gaussian separability criteria based on Positive Partial Transposition fails to reveal entanglement in the system. Following this, we have demonstrated that by performing tests based on a non-Gaussian generalization of the PPT criterion we can verify the existence of entanglement for any combination of GQW energy levels. 

Finally, we have examined the effects of considering particles at different gravitational potentials and shown that the entanglement of states persists even if one of the parts of the system moves to a weaker gravitational field.

\begin{acknowledgements}
This work is supported by the COST action MP1405. The work of CB is supported by the European Research Council (ERC-2010-AdG Grant 267841).
 \end{acknowledgements}

\end{document}